\documentclass{appolb}

\usepackage{graphicx}
\usepackage{subfigure}
\usepackage{cite}

\begin{document}

\title{Application of chiral quarks to high-energy processes and lattice QCD%
\thanks{Presented by WB at {\em Excited QCD}, Zakopane, 8-14 February 2009}%
}

\author{Wojciech Broniowski
\address{The H. Niewodnicza\'nski Institute of Nuclear Physics PAN, PL-31342 Krak\'ow\\
and Institute of Physics, Jan Kochanowski University, PL-25406~Kielce, Poland}
\and
Enrique Ruiz Arriola
\address{Departamento de F\'{\i}sica At\'omica, Molecular y Nuclear, Universidad de Granada, E-18071 Granada, Spain}
}

\date{February 2009}

\maketitle
\begin{abstract}
Results of the chiral quark models for the soft matrix
elements involving pions and photons, relevant for high-energy processes, are reviewed. 
We discuss quantities related to the
generalized parton distributions of the pion: the parton distribution
functions, the parton distribution amplitudes, and the generalized
form factors. The model predictions are compared
to the data or lattice simulations, with good
agreement. The QCD evolution from the low quark model
scale up to the experimental scales is a crucial ingredient of the approach.
\end{abstract}

 
\bigskip
 
The low-energy properties of the pion are dominated by the spontaneous
breakdown of the chiral symmetry, which is a key dynamical factor. It
allows to model the soft matrix element in a genuinely dynamical way
\cite{Davidson:1994uv,Dorokhov:1998up,Polyakov:1998td,Polyakov:1999gs,Dorokhov:2000gu,Anikin:2000th,Anikin:2000sb,RuizArriola:2001rr,Davidson:2001cc,RuizArriola:2002bp,RuizArriola:2002wr,Praszalowicz:2002ct,Tiburzi:2002kr,Tiburzi:2002tq,Theussl:2002xp,Broniowski:2003rp,Praszalowicz:2003pr,Bzdak:2003qe,Noguera:2005cc,Tiburzi:2005nj,Broniowski:2007fs,Courtoy:2007vy,Courtoy:2008af,Kotko:2008gy}.
There are two basic elements in our analysis: the low-energy dynamical
quark model itself, and the QCD {\em evolution}, bringing the
predictions from the low quark-model scale to higher scales of the
experiment or lattice data. This talk is based on 
Refs.~\cite{Broniowski:2007si,Broniowski:2008hx}, where the details
can be found.

The theoretical framework is conveniently set by the 
Generalized Parton Distributions (GPDs) 
\cite{Ji:1998pc,Radyushkin:2000uy,Goeke:2001tz,Bakulev:2000eb,Diehl:2003ny,Ji:2004gf,Belitsky:2005qn,Feldmann:2007zz,Boffi:2007yc}.
For the case of the pion, considered here, the GPD for the non-singlet channel is defined as 
\begin{eqnarray}
&& \epsilon_{3ab}\,{\cal H}^{q,{\rm NS}}(x,\zeta,t) \!=\!\! \int
\frac{dz^-}{4\pi} e^{i x p^+ z^-}\!\!\!\! \left . \langle \pi^b (p') | \bar \psi (0)
\gamma^+ \psi (z) \, \tau_3 | \pi^a (p) \rangle \right |_{z^+=0,z^\perp=0}, \nonumber 
\end{eqnarray}
with similar expressions for the singlet quarks and gluons. We have omitted the gauge link operators, absent in the 
light-cone gauge.
The kinematics is determined by $p'=p+q$, $p^2=p'^2=m_\pi^2$, $q^2=-2p\cdot q=t$, and $\zeta = q^+/p^+$, which denotes the momentum transfer 
passed along the light cone. 
Formal properties of GPDs can be compactly written in the symmetric notation involving
$\xi= \frac{\zeta}{2 - \zeta}$, $X = \frac{x - \zeta/2}{1 - \zeta/2}$, where one has
\begin{eqnarray}
H^{I=0}(X,\xi,t)=-H^{I=0}(-X,\xi,t), \; H^{I=1}(X,\xi,t)=H^{I=1}(-X,\xi,t). \nonumber
\end{eqnarray} 
For $X \ge 0$ one finds
%\begin{eqnarray}
${\cal H}^{I=0,1}(X,0,0) = q^{S, NS}(X)$,
%\end{eqnarray}
where $q(x)^i$ denote the parton distribution functions (PDFs).
The following {\em sum rules} hold:
\begin{eqnarray}
\int_{-1}^1 \!\!\!\!\! dX\, {H}^{I=1}(X,\xi,t) = 2 F_V(t), 
\; \int_{-1}^1 \!\!\!\!\! dX\,X \, {H}^{I=0}(X,\xi,t) = 2\theta_2(t)-2\xi^2 \theta_1(t), \nonumber
\end{eqnarray}
where $F_V(t)$ denotes the {\em electromagnetic form factor}, while $\theta_1(t)$ and $\theta_2(t)$
are the {\em gravitational form factors} \cite{Donoghue:1991qv}.
Other important formal properties are the {\em polynomiality} conditions~\cite{Ji:1998pc}, 
the {\em positivity bounds}~\cite{Pire:1998nw,Pobylitsa:2001nt}, and a {low-energy theorem}~\cite{Polyakov:1998ze} 
relating the GPD to the  pion {\em distribution amplitude}.
We stress that all these required properties are 
satisfied in our calculation \cite{Broniowski:2007si}.

\begin{figure}[tb]
\begin{center}
\subfigure{\includegraphics[width=0.48\textwidth]{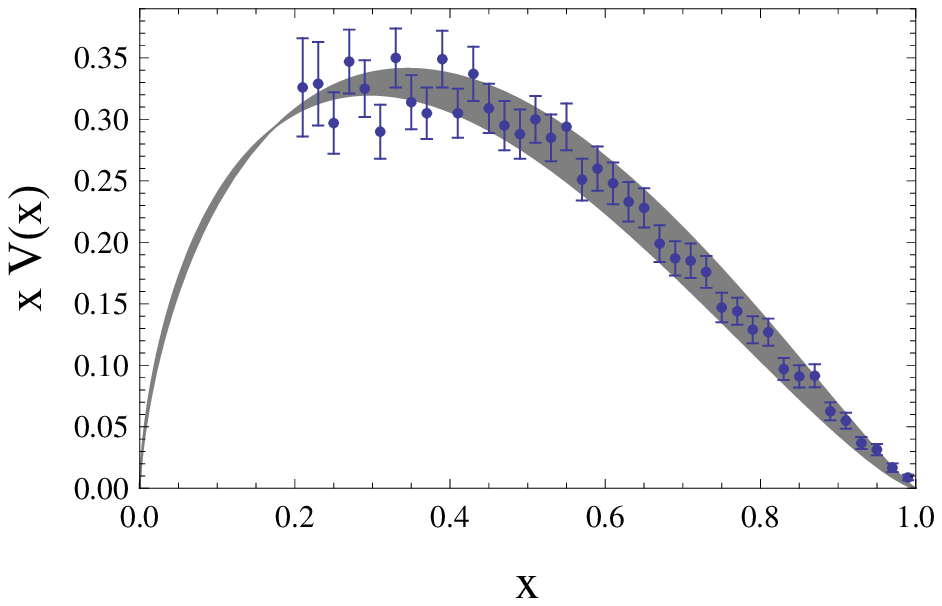}} \hfill
\subfigure{\includegraphics[width=0.48\textwidth]{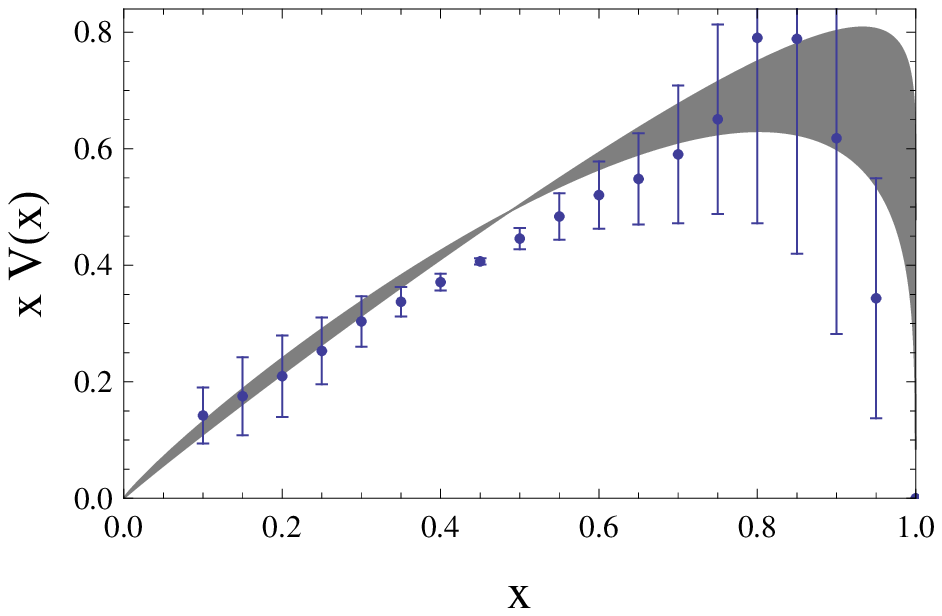}}
\end{center}
\vspace{-3mm}
\caption{Left: chiral quark model prediction for the valence
PDF of the pion, evolved to the scale of $Q=4$~GeV (band). The
width of the band indicates the uncertainty in the initial scale
$Q_0$. The data points come from the E615 experiment \cite{Conway:1989fs}. Right: same, evolved to the scale $Q=0.35~{\rm
GeV}$ and compared to the data from the transverse lattice
calculations of Ref.~\cite{Dalley:2002nj} \label{fig:pdf}}
\end{figure}

With $\zeta=t=0$, the GPDs becomes the usual PDFs. 
In the Nambu--Jona-Lasinio model~\cite{Davidson:1994uv} ${q(x)=1}$. This result holds at the low-energy 
quark-model scale, which needs to be determined.
At this scale the quarks are the only 
degrees of freedom. Thus, all observables are saturated with the quark contribution. In particular, this holds for the momentum 
sum rule. From experiment, the momentum fraction carried by the valence quarks is~\cite{Sutton:1991ay,Gluck:1999xe}
\begin{eqnarray} 
\langle x \rangle_v = 0.47(2) {\rm ~~at~~} Q^2 = 4~{\rm GeV}^2. \nonumber
\end{eqnarray} 
We evolve this value backward with the LO DGLAP equations down to the scale where the quarks carry 100\% of 
the momentum, $\langle x \rangle_v = 1$. 
This procedure yields the {\em quark model scale}
\begin{eqnarray}
{Q_0 = 313_{-10}^{+20} {\rm~MeV}}, \nonumber
\end{eqnarray} 
where the range reflects the uncertainty in $\langle x \rangle_v$.

The results of this method for the non-singlet PDF of the pion are
shown in the left panel of Fig.~\ref{fig:pdf}. We have evolved the
quark model result from the scale $Q_0$ up to the scale $Q=4$~GeV
corresponding to the E615 experiment \cite{Conway:1989fs}.  We notice
a very good agreement.  In the right panel we present the same
quantity evolved to the scale of $Q=350$~MeV and confronted with the
{\em transverse lattice} data \cite{Dalley:2002nj}, designed to work
at low-energy scales. Again, the agreement is remarkable.

\begin{figure}[tb]
\begin{center}
\subfigure{\includegraphics[angle=0,width=0.48\textwidth]{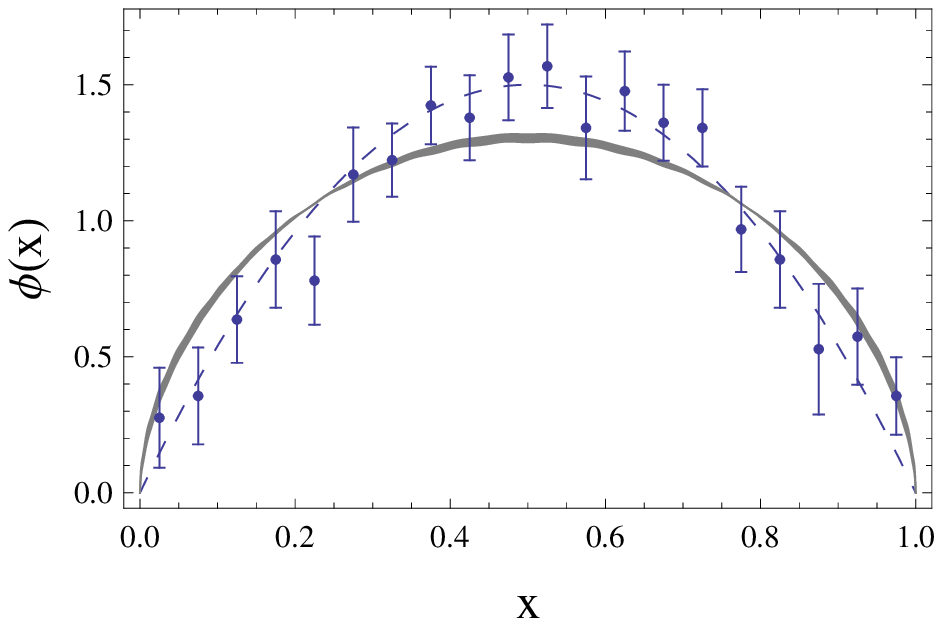}} \hfill
\subfigure{\includegraphics[angle=0,width=0.48\textwidth]{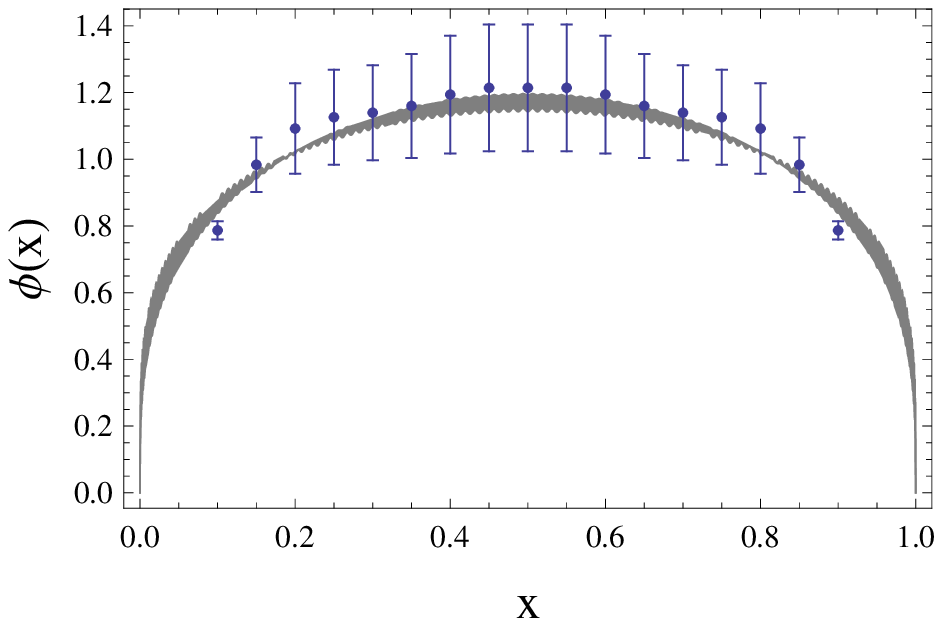}}
\end{center}
\vspace{-3mm}
\caption{Left: chiral quark model prediction for the pion DA evolved
  to $Q=2~{\rm GeV}$ (band) and compared to the E791
  data~\cite{Aitala:2000hb}. The width of the band indicates the
  uncertainty in the initial scale $Q_0$. The dashed line shows the
  asymptotic form $\phi(x,\infty)=6x(1-x)$. Right: same, evolved to
  $Q=0.5 {\rm GeV}$ and compared to the transverse lattice
  data~\cite{Dalley:2002nj}.}
\label{fig:pda}
\end{figure}

Next, we look at the DA. Here the evolution is carried out with the LO ERBL
equations. The results, displayed in Fig.~\ref{fig:pda}, again are in
fair agreement with the data, especially for the lattice case shown in the
right panel. 

\begin{figure}[tb]
\begin{center}
\subfigure{\includegraphics[width=.48\textwidth]{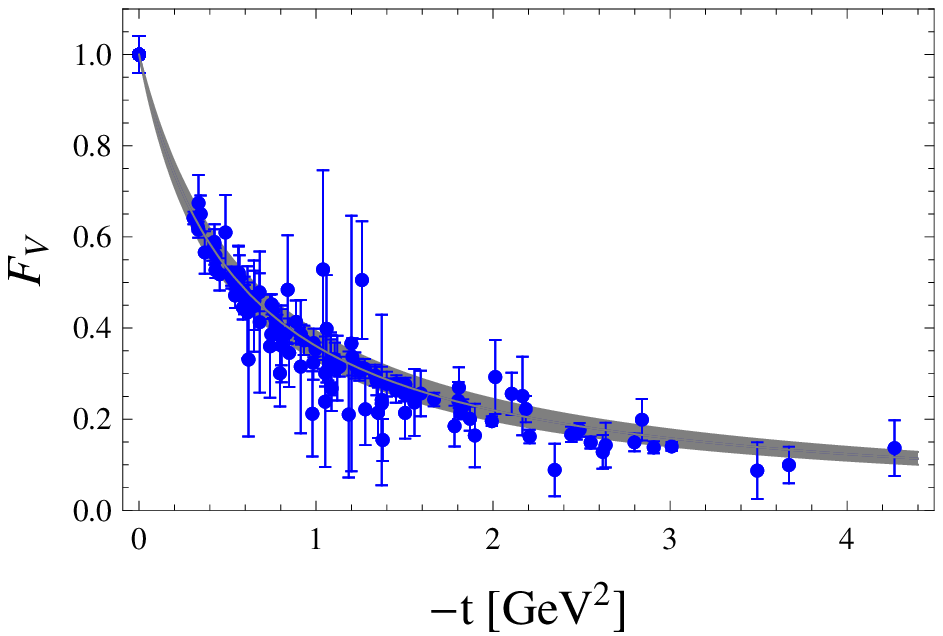}} \hfill
\subfigure{\includegraphics[width=.48\textwidth]{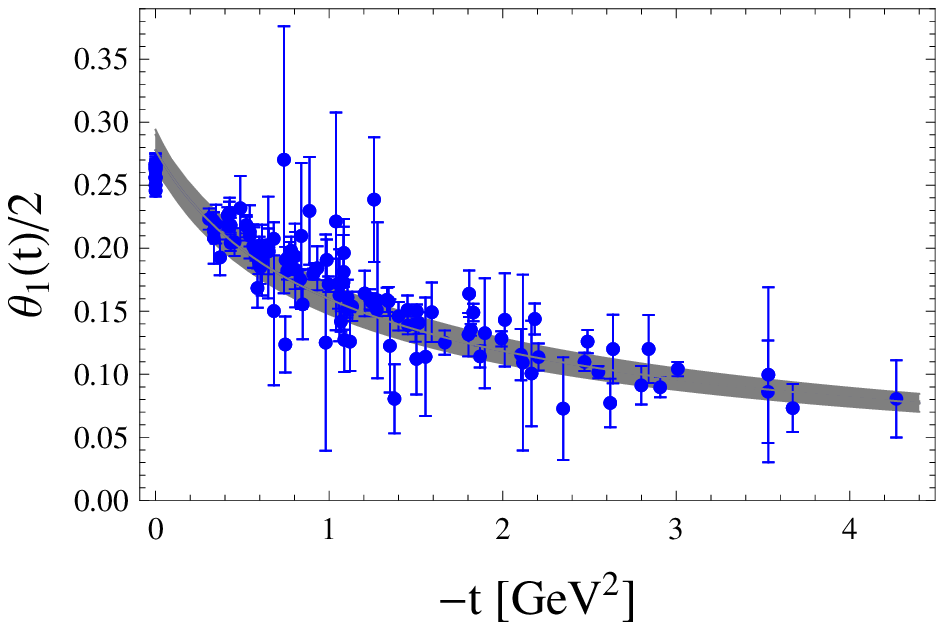}} 
\end{center}
\vspace{-3mm}
\caption{Left: the electromagnetic form factor. Right: the
  quark part of the gravitational form factor, $\theta_1(t)/2$, both computed in the Spectral Quark Model and 
compared to the lattice data from
  Ref.~\cite{Brommel:PhD}. The band around the model curves indicates
  the uncertainty in the quark momentum fraction. \label{fig:ff}}
\end{figure}

In Ref.~\cite{Broniowski:2007si} we provide formulas for 
the GPDs in the NJL model and in the Spectral Quark Model
\cite{RuizArriola:2003bs}.  These expressions have a rather non-trivial
structure, {\em not exhibiting factorization} in
the $t$ and $x$ variables, while satisfying all the formal requirements
mentioned above.  Since there is no data for the
full kinematic range for the GPDs, we only present the results for the
generalized form factors, for which there is recent information from the lattice QCD
\cite{Brommel:PhD,Brommel:2005ee}.  The vector
form factor and the quark part of the gravitational form factor of the
pion, obtained in the Spectral Quark Model, are compared to these lattice data in
Fig.~\ref{fig:ff}.  We note a very good agreement. In the Spectral
Quark Model the expressions are particularly simple,
\begin{eqnarray}
F_V^{\rm SQM}(t)=\frac{m_\rho^2}{m_\rho^2-t}, \;\;\;\; \theta_{1,2}^{\rm SQM}(t)/\theta_{1,2}^{\rm SQM}(0)=\frac{m_\rho^2}{t} \log
\left ( \frac{m_\rho^2}{m_\rho^2-t} \right ). \nonumber 
\end{eqnarray}
We note the longer tail of the gravitational form factor in the momentum space, meaning a more compact distribution 
in the coordinate space. Explicitly, we find a quark-model formula
$2 \langle r^2 \rangle_\theta = \langle r^2 \rangle_V$.

Finally, we compare our model values for the higher-order form factors at
$t=0$ to the lattice data provided in Sec.~7 of
Ref.~\cite{Brommel:PhD}, given below in parenthesis.  
After the evolution to the lattice scale of $Q=2$~GeV we find
\begin{eqnarray}
&&\langle x \rangle = 0.28\pm 0.02 \;(0.271\pm 0.016), \nonumber \\
&& \langle x^2 \rangle = 0.10\pm 0.02 \;(0.128\pm 0.018), \nonumber \\
&& \langle x^3 \rangle = 0.06\pm 0.01 \;(0.074\pm 0.027).  \nonumber
\end{eqnarray}
The model error bars come from the uncertainty of $Q_0$. We note that
the model predictions fall within the error bars. It should be
mentioned that while high energies are needed to test 
leading twist parton distributions, transverse and
Euclidean lattices can directly evaluate them {\it per se}. In~\cite{Broniowski:evol} we provide a
handy way of undertaking evolution for generalized form factors which
are currently becoming directly available on Euclidean lattices.

To briefly summarize, the chiral quark models supplied with the QCD evolution 
work well for a wide variety of quantities related to the GPDs of the pion and 
provide valuable insight into the non-perturbative dynamics behind the soft matrix elements.

\bigskip
Supported in part by the Polish Ministry of Science and Higher Education, grants N202~034~32/0918 and N202~249235, Spanish DGI and
FEDER funds with grant  FIS2008-01143/FIS, Junta de Andaluc{\'\i}a grant FQM225-05, and the EU Integrated Infrastructure Initiative
Hadron Physics Project, contract RII3-CT-2004-506078.

%\bibliography{gff}
%\bibliographystyle{h-elsevier}

\end{document}